%% file: iclr2024_conference.tex
\documentclass{article} 
\usepackage{iclr2024_conference,times}
\usepackage{booktabs, multirow, makecell,longtable}

\input{math_commands.tex}

\usepackage{hyperref}
\usepackage{url}
\usepackage{graphicx}
\usepackage{enumitem}

\title{DrugAssist: A Large Language Model for Molecule Optimization}

\iclrfinalcopy

\author{Geyan Ye,$^{1\dagger}$ ~~~~Xibao Cai,$^{2\dagger}$ ~~~~Houtim Lai,$^{1}$ ~~~~Xing Wang,$^{1}$ ~~~~Junhong Huang,$^{1}$ \\ \bf Longyue Wang,$^{1}$\thanks{~Corresponding Author, $^\dagger$ Equal Contribution.} ~~~~Wei Liu$^{1}$~~~~\&~~~~Xiangxiang Zeng$^{2}$\\
$^{1}$Tencent AI Lab ~~~~~ $^{2}$Department of Computer Science, Hunan University\\
\texttt{\{blazerye,vinnylywang,topliu\}@tencent.com}\\
\texttt{\{dalecai,xzeng\}@hnu.edu.cn}
}

\begin{document}

\maketitle

\begin{abstract}
Recently, the impressive performance of large language models (LLMs) on a wide range of tasks has attracted an increasing number of attempts to apply LLMs in drug discovery. However, molecule optimization, a critical task in the drug discovery pipeline, is currently an area that has seen little involvement from LLMs. Most of existing approaches focus solely on capturing the underlying patterns in chemical structures provided by the data, without taking advantage of expert feedback. These non-interactive approaches overlook the fact that the drug discovery process is actually one that requires the integration of expert experience and iterative refinement. To address this gap, we propose DrugAssist, an interactive molecule optimization model which performs optimization through human-machine dialogue by leveraging LLM’s strong interactivity and generalizability. DrugAssist has achieved leading results in both single and multiple property optimization, simultaneously showcasing immense potential in transferability and iterative optimization. In addition, we publicly release a large instruction-based dataset called ``MolOpt-Instructions'' for fine-tuning language models on molecule optimization tasks. We have made our code and data publicly available at \url{https://github.com/blazerye/DrugAssist}, which we hope to pave the way for future research in LLMs' application for drug discovery.
\end{abstract}

\begin{figure}[h]
    \centering
    \includegraphics[width=0.83\textwidth]{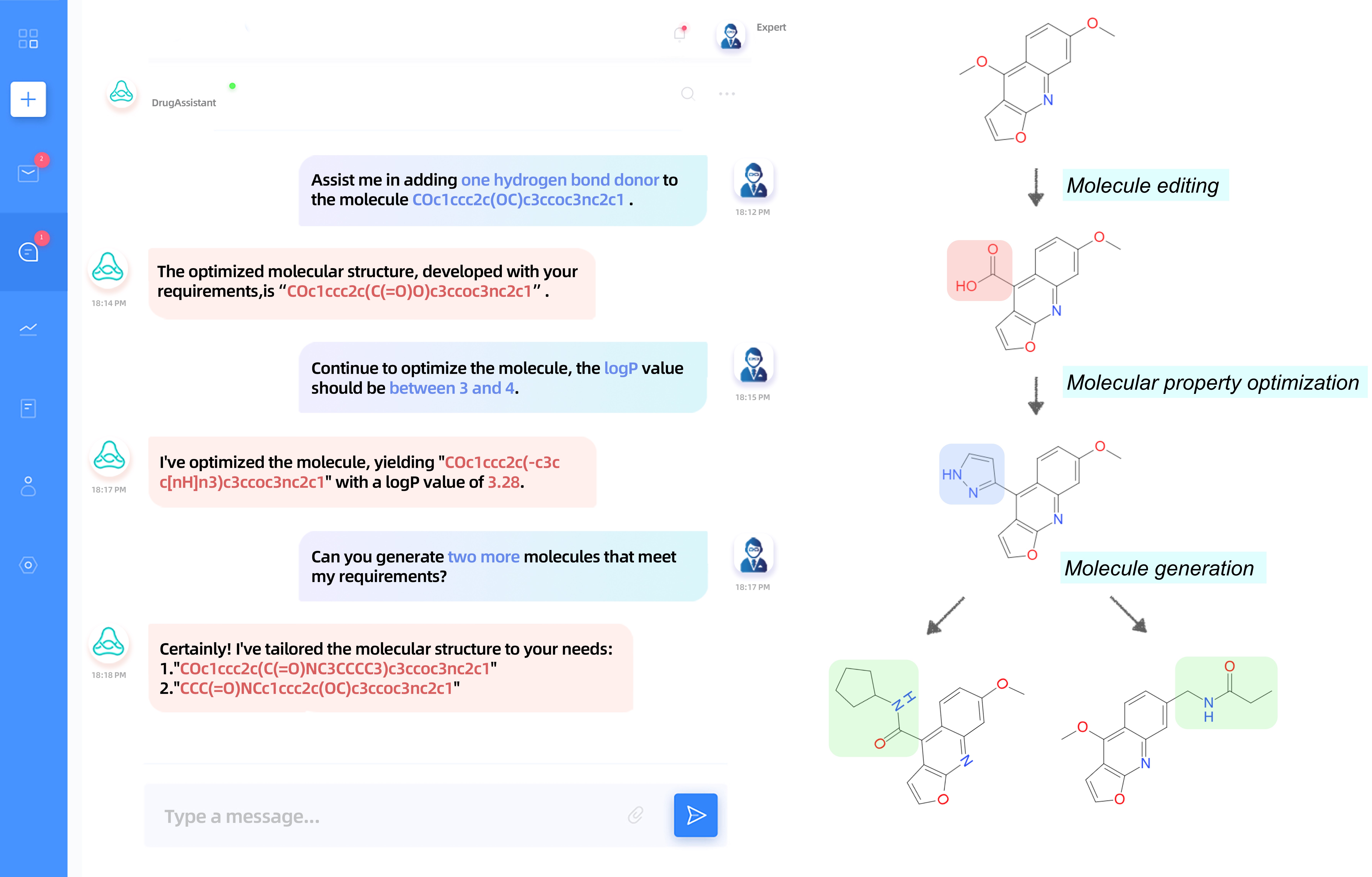}
    \caption{The illustration of our proposed DrugAssist model framework, which focus on optimizing molecules through human-machine dialogue.}
    \label{fig:abs_fig}
\end{figure}

\clearpage

\section{Introduction}

Recently, generative artificial intelligence has made remarkable strides in the field of natural language processing (NLP), particularly with the advent of Large Language Models (LLMs) such as GPT (Generative Pre-trained Transformer) \citep{radford2019language}. These models have demonstrated impressive capabilities in a wide range of tasks, extending far beyond everyday communication and question-answering scenarios. Researchers have increasingly recognized the potential of these  models in addressing complex and diverse problems across various domains, prompting interests in their applications within professional fields. 

In recent years, there has been an increasing number of attempts to apply conversational LLMs in the field of drug discovery \citep{yunxiang2023chatdoctor,han2023medalpaca,wu2023pmc,luo2023biomedgpt,zeng2023interactive,liu2023chatgpt}. However, molecule optimization, a critical task in the drug discovery pipeline, is currently an area that has seen little involvement from LLMs. Existing approaches can be broadly categorized into two main types. The first type represents molecules as sequences (commonly SMILES strings) and generates an optimized molecular sequence by learning from the input data. The second type of approach represents molecules as graphs and formulates molecule optimization as a graph-to-graph translation problem \citep{he2022transformer}. One of the major issues with these approaches is the lack of interactivity. They focus solely on capturing the underlying patterns in chemical structures provided by the data, without taking advantage of the invaluable expert experience and feedback. In contrast, the drug discovery pipeline involves iterative refining processes that entail conversations with domain experts to incorporate their feedback, ultimately achieving the desired outcome \citep{liu2023chatgpt}.

In light of the advancements in powerful LLMs, our work aims to leverage their strong interactivity and generalizability for molecule optimization. To the best of our knowledge, there are currently no molecule optimization models that focus on human-machine interaction. We summarize main contributions of this work as follows:
\begin{itemize}[leftmargin=*,topsep=0.1em,itemsep=0.1em,parsep=0.1em]
\item To facilitate future research, we publicly release a large instruction-based dataset called ``MolOpt-Instructions'' for fine-tuning language models on molecule optimization tasks. The dataset contains an adequate amount of data, ensuring both similarity constraints and a substantial difference in properties between molecules.

\item We propose DrugAssist, an interactive molecule optimization model fine-tuned on Llama2-7B-Chat, which performs optimization through human-machine dialogue. By enabling multi-turn conversations, domain experts can guide the model in further optimizing initially generated molecules with imperfections.

\item Compared to traditional molecular optimization approaches \citep{he2021molecular,jin2020hierarchical} and LLM-based implementations \citep{liu2023chatgpt,luo2023biomedgpt}, DrugAssist has consistently achieved leading results in multi-property optimization, which is a less frequently addressed and more challenging task in molecule optimization. Moreover, our optimization objectives include maintaining optimized molecular property values within a given range. DrugAssist continues to demonstrate impressive performance in this category of tasks, which are more aligned with real-world requirements compared to most studies that solely focus on increasing or decreasing property values.
\end{itemize}

\section{Related Work}
\label{related work}

\subsection{Traditional approaches in molecule optimization} 

Based on the different representations of molecules, we can divide these models into two categories: sequence-based and graph-based.

\paragraph{Sequence-based}  Most of these methods utilize SMILES (Simplified Molecular-Input Line-Entry System) string as the molecular representation. They view molecule optimization as machine translation problem in natural language processing (NLP), where a text is translated from one language to another \citep{he2021molecular}. Similar to translation tasks in NLP, the conversion between molecules encoded in SMILES in molecular optimization tasks can also be seen as a transformation between ``languages''. The main architectures for this category of models include recurrent neural networks (RNNs) \citep{gupta2018generative,bjerrum2017molecular,segler2018generating}, variational autoencoders (VAEs) \citep{jin2018junction,jin2018learning,dai2018syntax,liu2018constrained,simonovsky2018graphvae}, and Transformer \citep{he2021molecular,he2022transformer}. Meanwhile, reinforcement learning \citep{olivecrona2017molecular,putin2018reinforced}, adversarial training \citep{kadurin2017drugan}, and transfer learning \citep{segler2018generating} serve as typical optimization techniques. Considering the significant progress has made in the field of LLMs in recent years, we believe that these sequence-based molecule optimization methods have great potential for further exploration.

\paragraph{Graph-based} These methods typically use graph to represent molecules and directly generate molecule graphs. VAEs are also very popular in molecular graph generation. \citet{jin2018junction} decomposed a molecular graph into a junction tree of chemical substructures. Then, they employed a junction tree VAE (JT-VAE) to generate molecules with improved properties by applying gradient ascent over the learned latent space. Subsequent works have derived many variants based on JT-VAE, such as VJTNN \citep{jin2018learning}, HierG2G \citep{jin2020hierarchical}, etc. In comparison to sequence-based methods, a notable distinction is that most graph-based approaches, such as JT-VAE, can consistently generate valid molecules due to the validation checks performed at each step of the generation process.

Despite the achievements of the aforementioned methods in the field of molecule optimization, we believe that they still have some shortcomings that need to be addressed:

\begin{itemize}[leftmargin=*,topsep=0.1em,itemsep=0.1em,parsep=0.1em]
\item Most of the existing works focus on optimizing a single property of molecules, while there are few that simultaneously optimize multiple properties, which is a more common requirement in real life. Moreover, in most works, the optimization goal is to maximize the difference in properties between the optimized and original molecules while satisfying a certain similarity constraint. Alongside this, it's worth noting that in real-life situations, there is often a need for the property value of the optimized molecule to fall within a specific range, an aspect that has received little attention in existing research.

\item Most methods suffer from catastrophic forgetting when the optimization task is changed. For example, a model that performs well in optimizing QED values of molecules needs to be retrained on a dataset containing logP property before being used to optimize logP values. This approach not only incurs additional costs but also suffers from a lack of sufficient experimental data to facilitate training for some molecular properties, such as ADMET.

\item To the best of our knowledge, no existing studies have focused on the interactivity of these molecular optimization models. Interactive models facilitate effective communication between domain experts and artificial intelligence models. Experts can conveniently provide feedback and suggestions to the model in the form of natural language, and the model can also obtain real-time access to expert experience related to specific problems. However, existing approaches struggle to efficiently utilize these valuable expert experiences and feedback.

\end{itemize}

\subsection{LLMs in biomedical domain}  

In recent years, there has been an increasing number of attempts to apply LLMs in the field of biomedicine. The majority of research efforts are centered around QA (question-answering) tasks, such as Chatdoctor \citep{yunxiang2023chatdoctor}, Med-Alpaca \citep{han2023medalpaca}, PMC-LLaMA \citep{wu2023pmc} that focus on medical QA, and BioMedGPT \citep{luo2023biomedgpt} on molecule/protein QA. There is relatively much less work focused on addressing practical tasks within the drug discovery domain. ChatMol \citep{zeng2023interactive} is employed for conversational molecular design, specifically, it can accomplish two tasks: molecule understanding and molecule generation, which involve bidirectional conversion between molecular descriptions and SMILES strings of molecules. ChatDrug \citep{liu2023chatgpt} is a framework to facilitate drug editing using LLMs. Specifically, following the ChatDrug workflow, users can obtain carefully crafted prompts that assist in obtaining suggestions on drug editing tasks from general LLMs, such as ChatGPT.

\section{Methods}
\label{methods}

Our methodology incorporates two primary components: the construction of our MolOpt-Instructions dataset and the subsequent instruction tuning of Llama2-7B-chat model.

\subsection{Construction of MolOpt-Instructions Dataset}

Most of datasets currently used for molecule optimization are in the form of ``molecule-molecule pairs'', which cannot be directly used to train language models like Llama. Although \citet{fang2023mol} introduce a comprehensive instruction dataset specifically designed for training Large Language Models in the biomedical domain, it does not cover tasks associated with molecule optimization. Additionally, in some popular benchmark dataset \citep{jin2018learning}, the molecule pairs which are relatively few in number, only satisfy similarity constraints, and the difference in properties between the molecules within the same pair is not sufficiently significant. To tackle these issues, we construct instruction-based datasets called ``MolOpt-Instructions'' for fine-tuning language models on molecule optimization tasks. 
It contains an adequate amount of data, ensuring both similarity constraints and a substantial difference in properties between molecules.

\paragraph{Overview and Statistics}

MolOpt-Instructions consists over one million molecule pairs. Currently, it includes six types of molecular properties, namely Solubility, BBBP (Blood-Brain Barrier Penetration), hERG (Human Ether-a-go-go-Related Gene) inhibition, QED (Quantitative Estimate of Drug-likeness) and the number of hydrogen bond donor and acceptor, with detailed information provided in Table ~\ref{tab:cmp-dataset}. 

\begin{table}[h]
    \centering
    \begin{tabular}{cccc}
        \toprule
         \textbf{Unique pairs} & \textbf{Unique molecules} & \textbf{Similarity} & \textbf{LogP difference}  \\
        \midrule
         1,029,949 & 1,595,839 & $0.69\pm0.06$ & $2.82\pm0.31$ \\

        \bottomrule
    \end{tabular}
    \caption{Statistics of our proposed MolOpt-Instructions dataset. It contains an adequate amount of data, ensuring both similarity constraints and a substantial difference in properties between molecules.}
\label{tab:cmp-dataset}
\end{table}

\paragraph{Data Construction}

\begin{figure}[t]
    \centering
    \includegraphics[width=1\textwidth]{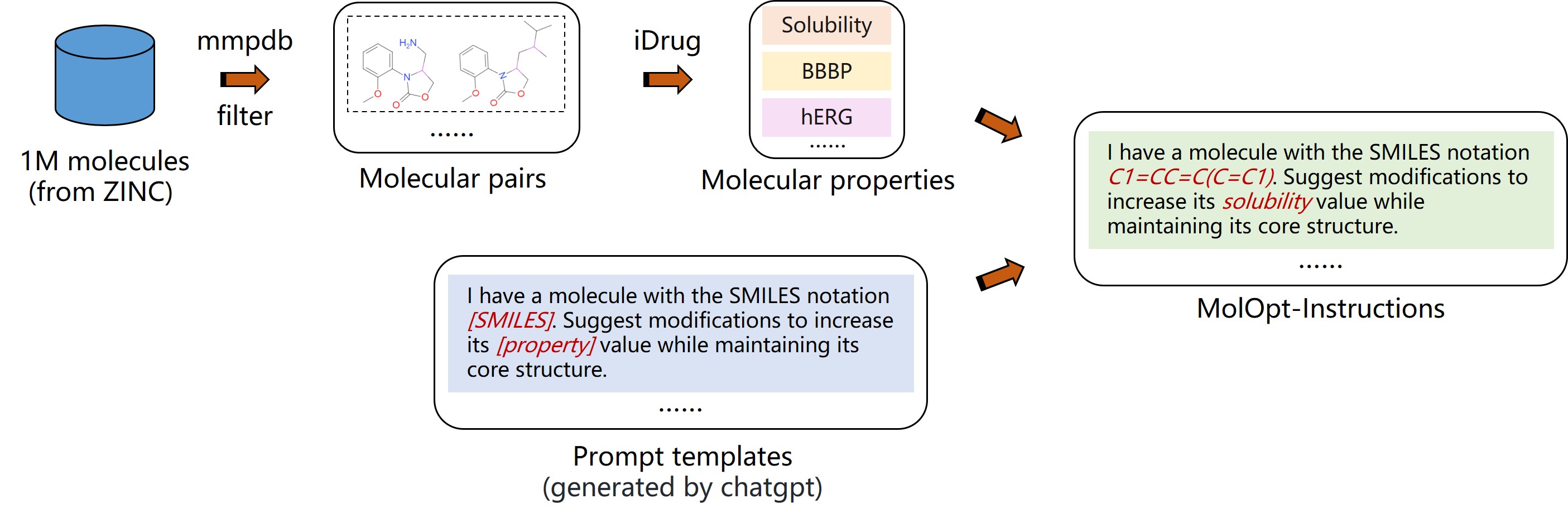}
    \caption{The workflow of data construction of MolOpt-Instructions. First, we randomly picked one million molecules from the ZINC dataset. Then, we used mmpdb \citep{dalke2018mmpdb} to generate similar pairs based on these molecules and selected the molecular pairs that met our requirements from these candidates. Once we identified the suitable molecular pairs, we proceeded to calculate their property values using iDrug \citep{idrug}. After obtaining these pairs and their corresponding property values, we asked ChatGPT to suggest a variety of instructions and manually refine them for the molecule optimization tasks.}
    \label{fig:case_intro}
\end{figure}

The workflow of the data construction is shown in Figure ~\ref{fig:case_intro}. To begin with, we randomly selected one million molecules from ZINC database \citep{irwin2005zinc}. Then, we used mmpdb \citep{dalke2018mmpdb} to construct a database from these molecules and generate similar pairs. Mmpdb, an open-source Matched Molecular Pair (MMP) platform, generates MMPs through Matched Molecular Pair Analysis (MMPA). In essence, a MMP consists of two molecules that differ by a defined structural transformation, resulting in highly similar molecular structures within the pairs generated by mmpdb. Following this, we selected the molecular pairs that met our requirements from these candidates. Our selection criteria are as follows: \textit{the similarity between each pair of molecules should be greater than 0.65, and the difference in logP should be greater than 2.5.} Once we identified the suitable molecular pairs, we proceeded to calculate their property values using iDrug, an AI-driven drug discovery platform developed by Tencent \citep{idrug}. To make the data more balanced, we maintain a roughly 1:1 ratio of increased to decreased property values for target molecules relative to source molecules by swapping the source and target molecules of some pairs. The rationale behind choosing the difference in logP as a screening criterion lies in its close relation to various aspects of a molecule's biological activity and pharmacokinetics. After obtaining these pairs and their corresponding property values, we asked ChatGPT to suggest a variety of instructions and manually refine them for the molecule optimization tasks. 

We designed three types of optimization tasks: the first category requires only an increase or decrease in the given property value; the second category adds a threshold requirement for the increase or decrease; and the third category requires the optimized property value to be within a given range. In Table~\ref{tab:prompt_ex}, we show an example of an instruction for each of these three categories. All instructions in the dataset can be found in \url{https://github.com/blazerye/DrugAssist}.

\begin{table}[t]
\centering
\begin{tabular}{@{}cp{11cm}@{}}
\toprule
\textbf{Task category} & \textbf{Example prompt} \\
\midrule
loose & I have a molecule with the SMILES string [SMILES]. Suggest modifications to increase its [property] value while maintaining its core structure.
 \\
\midrule
strict & I have a molecule with the SMILES string [SMILES]. Suggest modifications to increase its [property] value by at least [threshold] compared to the pre-optimized value while maintaining its core structure.
 \\
\midrule
range & Here is a molecule represented by the SMILES string [SMILES]. Provide me with an optimized version that has a [property] value between [lower bound] and [upper bound]. The output molecule should be similar to the input molecule.
 \\

\bottomrule
\end{tabular}
\caption{Examples of prompts for optimization tasks with three different goals - loose, strict and range. ``[SMILES]'' represents the SMILES string for the molecule.}
\label{tab:prompt_ex}
\end{table}

Different from several widely used molecule optimization datasets, our optimization tasks are not just vaguely asking to ``optimize the given molecule'', but also have range requirements, making them more closely aligned with real-world scenarios.

\paragraph{Analysis and Discussion}

To ensure the diversity of molecules in our dataset, we employ Murcko scaffold analysis to evaluate the chemical diversity of the source molecules randomly selected from ZINC database. The average molecules per scaffold is 2.95, and more than 93.7\% of the scaffolds contained no more than five molecules. The scaffold analysis indicates a high degree of structural diversity among the source molecules. Therefore, models developed using this dataset are expected to demonstrate robust prediction coverage for a broad spectrum of structurally diverse compounds. 

\begin{figure}[t]
    \centering
    \includegraphics[width=0.8\textwidth]{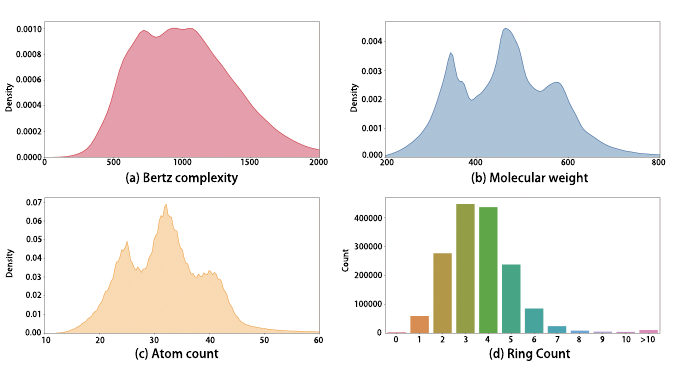}
    \caption{Distribution of structural properties of molecules within MolOpt-Instructions, illustrating the structural diversity of the molecules.}
    \label{fig:phy_fig}
\end{figure}

\begin{figure}[t]
    \centering
    \includegraphics[width=1\textwidth]{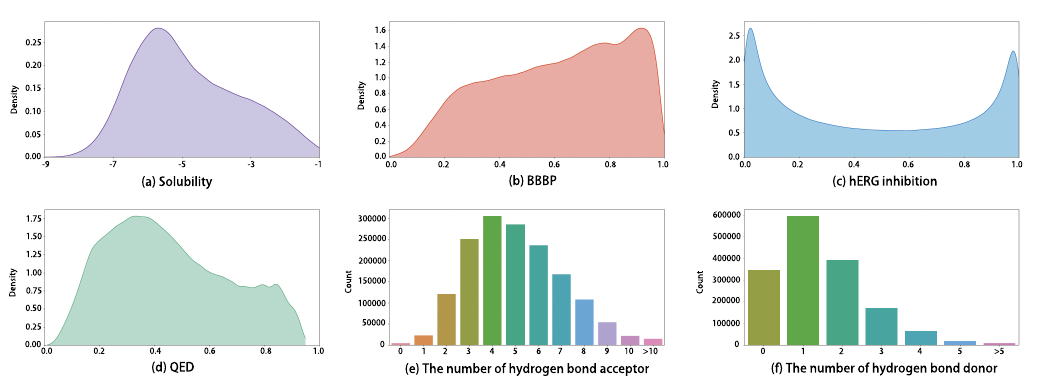}
    \caption{Distribution of ADMET-related properties of molecules within MolOpt-Instructions. Currently, MolOpt-Instructions covers six properties, namely Solubility, BBBP (Blood-Brain Barrier Penetration), hERG (Human Ether-a-go-go-Related Gene) inhibition, QED (Quantitative Estimate of Drug-likeness) and the number of hydrogen bond donor and acceptor. The distribution graph demonstrates the diversity of biochemical properties of molecules in our dataset.}
    \label{fig:admet_fig}
\end{figure}

Furthermore, we also plot distribution graphs for molecular structural and ADMET-related properties, as shown in Figure ~\ref{fig:phy_fig} and Figure ~\ref{fig:admet_fig}, respectively. For molecular structural properties, we focus on Bertz complexity, molecular weight, atom count and ring count. Bertz Complexity is a key parameter for assessing the structural complexity of a molecule, providing insights into its potential reactivity and stability. Molecular weight, a measure of a molecule's size, influences various physical and chemical properties, including solubility, volatility, and reaction kinetics. Atom count, indicative of the molecule's size and complexity, impacts its stability and potential intermolecular interactions. Ring count, a measure of cyclic structures within a molecule, informs about its structural rigidity, conformational flexibility, and possible biological activity. These graphs provide a more intuitive visualization of the diversity in the physical structure and biochemical properties of molecules in MolOpt-Instructions.

\subsection{Instruction Tuning}
For LLMs to follow natural language instructions and complete real-world tasks, instruction tuning has been widely used for alignment \citep{ouyang2022training}. In this process, the LLM is fine-tuned on a collection of tasks, which are defined through a set of instructions. 

Our work follow the similar approach performing instruction tuning on Llama2-7B-Chat using our MolOpt-Instructions dataset. Formally, we define the text input as a sequence of tokens, \(U = \{u_{1}, u_{2}, \dots, u_{N}\}\), where each \(u_i\) is a text token and \(N\) is the total sequence length. At the stage of instruction fine-tuning, the sequence \(U\) is further split into two parts, instruction  \(I\) and response \(R\). The training objective is to minimize the negative log-likelihood over the response \(R\) with respect to trainable parameters $\theta$ as follows: $ L(R;\theta) = -\sum_{u_i\in R}{\log \varPhi \left( u_i\middle| u_{<i},I \right)}$.

\begin{figure}[t]
    \centering
    \includegraphics[width=0.7\textwidth]{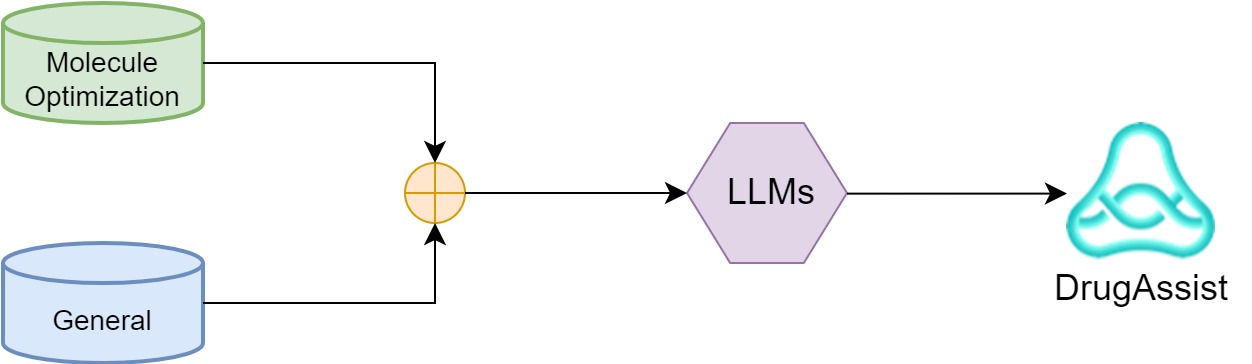}
    \caption{The illustration of multi-task learning strategy. We apply instruction tuning by directly combining different data sources (general knowledge and molecule optimization), effectively mitigating catastrophic forgetting during the fine-tuning stage.
}
    \label{fig:multitask_fig}
\end{figure}

\textbf{Multi-task learning} In instruction tuning, the occurrence of catastrophic forgetting is a common phenomenon in pre-trained language models \citep{de2021continual,dong2023abilities}. Ensuring that the model maintains high interactivity while optimizing molecules is one of our important objectives. To achieve this, we employ multi-task learning as our instruction tuning strategy. Specifically, the composition of our text input consists of two parts: (1) \textit{General knowledge}: such as everyday conversational question-answering data; (2) \textit{Domain-specific knowledge}: for our model, it pertains to molecule optimization. 
We mix these two types of data at a certain ratio. To achieve this ratio, we replicate the data from the less abundant category. Figure ~\ref{fig:multitask_fig} illustrates our instruction tuning strategy.

\section{Experiments}
We provide a comprehensive view of DrugAssist's performance on traditional molecule optimization tasks, as well as its capabilities in dialogue and interaction.

\subsection{Experimental Setup}

\paragraph{Models}

DrugAssist is a model fine-tuned from the Meta's Llama-2-7B-Chat model on over one million instruction-response demonstrations. We conduct a systematic comparison with the following sequence-based models:
\begin{itemize}[leftmargin=*,topsep=0.1em,itemsep=0.1em,parsep=0.1em]
\item \textit{\citet{he2021molecular}} utilized the SOTA machine translation models, the Seq2Seq with attention and the Transformer for molecule optimization tasks. 

\item \textit{ChatDrug} \citep{liu2018constrained} is a framework to facilitate the systematic investigation of drug editing using LLMs. For the molecule optimization tasks, they obtained results from ChatGPT (GPT-3.5-turbo) using carefully crafted prompts.

\item \textit{Llama2-7B-Chat} \citep{touvron2023llama} is a fine-tuned generative text model with 7 billion parameters, developed and publicly released by Meta. It outperforms open-source chat models on most benchmarks, and is on par with some popular closed-source models like ChatGPT and PaLM \citep{narang2022pathways}.

\item \textit{BioMedGPT-LM-7B} \citep{luo2023biomedgpt} is the first large generative language model based on Llama2 in the biomedical domain.
\end{itemize}

\paragraph{Training Details}
At instruction tuning stage, we train the model for 10 epochs with a batch size of 512. We use the AdamW optimizer, with $\beta$ =(0.9, 0.999) and a learning rate of 1e-4, without weight decay. Warm-up is executed over 3\% of the total training steps, followed by a cosine schedule for learning rate decay. Linear layers within the LLM utilize a LoRA rank of 64 and a LoRA alpha of 128. The model is trained on 8 NVIDIA Tesla A100-SXM4-40GB GPUs.

\paragraph{Dataset for Instruction Tuning}
To ensure our model maintains high interactivity while optimizing moluecules, we employ multi-task learning strategy introduced in Section 3.2 to construct training data. We utilize instruction data from two sources:
\begin{itemize}[leftmargin=*,topsep=0.1em,itemsep=0.1em,parsep=0.1em]
\item \textit{MolOpt-Instructions} We have provided a detailed introduction to this dataset in Section \ref{methods}.

\item \textit{Stanford Alpaca} In order to preserve the model's natural language dialogue capabilities and counteract the forgetting effect  during the supervised fine-tuning phase, we utilized the dataset employed for fine-tuning a 7B Llama model, which comprises 52k instruction-following data provided by Stanford.
\end{itemize}

Considering that the MolOpt-Instructions dataset contains significantly more data than the Stanford Alpaca dataset, we created the final dataset by replicating the Stanford Alpaca dataset five times and then mixing it with the Molopt-Instructions dataset. We divide the mixed data into training, validation, and test sets at a ratio of 0.9 : 0.05 : 0.05, respectively.

\subsection{Evaluation Methods}

\paragraph{Comparisons with Traditional Approaches} 
 
We compared DrugAssist with two molecule optimization models proposed by \citet{he2021molecular}. One of them employs a Seq2Seq with attention architecture (which we refer to as Mol-Seq2Seq), and the other uses a Transformer architecture (which we refer to as Mol-Transformer). Specifically, we compared the performance of these models in optimizing two properties: BBBP and Solubility. We calculated the success rates, validity, and average similarity between molecules before and after optimization, with the detailed definition of  ``success’’ summarized as follows:

\begin{itemize}[leftmargin=*,topsep=0.1em,itemsep=0.1em,parsep=0.1em]
\item \textit{Solubility}: We consider the optimization to be successful if the Solubility of the generated molecule falls within the given range. Specifically, we have divided the Solubility values into 10 intervals, each with a size of 1.

\item \textit{BBBP}: We consider the optimization to be successful if  the generated molecule's BBBP property type is correct. Specifically, we have categorized BBBP values into three groups: low, medium, and high, corresponding to the value ranges of 0-0.3, 0.3-0.7, and 0.7-1, respectively.

\end{itemize}

The prompt we used is ``\textit{Here is a molecule represented by the SMILES string [SMILES]. Provide me with an optimized version that has a molecular solubility value between [lower bound] and [upper bound] (unit: logarithm of mol/L), and change the blood-brain barrier penetration (BBBP) from [source category] to [target category]}''. We use this prompt to obtain results from DrugAssist in a single-turn dialogue manner.

\paragraph{Comparisons with LLMs} 

\begin{figure}[t]
    \centering
    \includegraphics[width=0.85\textwidth]{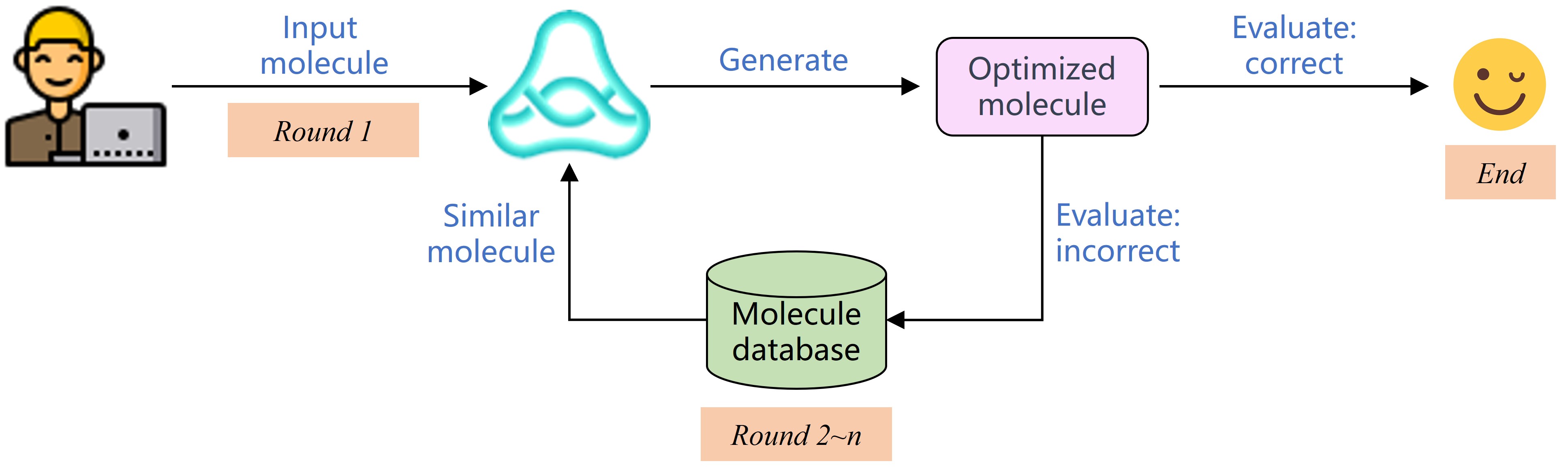}
    \caption{Multi-round optimization process proposed by \citet{liu2023chatgpt}. The model is initially provided with a source molecule and specific optimization criteria. Upon generating the optimized molecule, its property values are assessed. If it meets the requirements, the process terminates. If not, a  molecule most similar to the source molecule and fulfilling the criteria is retrieved from the database to guide further optimization. The process continues until the requirements is met or the predefined maximum number of iterations is reached.}
    \label{fig:multiround_fig}
\end{figure}

We compared DrugAssist with ChatDrug, Llama2-7B-Chat, and BioMedGPT-LM-7B. We randomly selected an additional 500 molecules from the ZINC dataset to serve as the testset for this experiment. Specifically, we compared the performance of these models on 16 tasks. Following the approach of \citet{liu2023chatgpt}, we employ multi-turn dialogues to enable LLMs to optimize molecules. We first propose optimization requirements, and if the model's output does not meet our requirements, we search the database for a molecule that meets the requirements and is most similar to the model's output, using it as a hint for the model to make modifications, until the requirements are met or the pre-set number of iterations is reached. Figure ~\ref{fig:multiround_fig} illustrates the optimization process.

We computed the success rate and validity for each task. We have adopted two sets of criteria for defining ``successful optimization'' - loose and strict. For the loose criteria, if the optimized molecular property is higher or lower than the pre-optimization property as required, we consider the optimization to be successful. For the strict criteria, except for the ``Solubility'', if the optimized molecular property is higher or lower than the pre-optimization property by a specified threshold, we consider the optimization to be successful. For the ``Solubility'', if the optimized molecular property value falls within the required range, we consider the optimization to be successful. Our range requirements are set as follows: Given a test molecule with Solubility value x, in the task of increasing the Solubility value, the optimized range requirement is [x+0.5, x+1.5]; in the task of decreasing the Solubility value, the optimized range requirement is [x-1.5, x-0.5]. The threshold settings for different properties in our experiments are shown in Table ~\ref{tab:threshold info}. Detailed prompt settings for each task can be found in Table ~\ref{tab:prompt_chatdrug} in Appendix. BBBP, Solubility, and hERG inhibition are predicted from iDrug, while the rest can be calculated deterministically using RDKit.

\begin{table}[t]
    \centering
    \begin{tabular}{lc}
        \toprule
        \textbf{Property} & \textbf{Threshold} \\
        \midrule
        QED  & 0.1  \\
        hydrogen bond acceptor  & 1 \\
        hydrogen bond donor  & 1 \\
        BBBP  & 0.1  \\
        hERG inhibition  & 0.1  \\

        \bottomrule
    \end{tabular}
    \caption{The threshold settings for different properties. For the strict criteria, except for the ``Solubility'', we consider the optimization to be successful only if the optimized molecular property is higher or lower than the pre-optimization property by the threshold shown in table.}
\label{tab:threshold info}
\end{table}

\subsection{Main Results}
In this section, we conduct a comprehensive and systematic comparison with aforementioned models.

\textbf{Comparisons with traditional approaches} Here we compared our model with \citet{he2021molecular}. The results are shown in Table ~\ref{tab:cmp-dmo}. Our model has achieved the highest success rates in both single-property and multi-property optimization while maintaining high validity and high similarity to the molecules to be optimized.

\begin{table}[t]
    \centering
    \begin{tabular}{lccccc}
        \toprule
        \textbf{Model} & \textbf{Solubility} & \textbf{BBBP} & \textbf{All} & \textbf{Valid rate} & \textbf{Similarity} \\
        \midrule
        \textbf{Mol-Seq2Seq}  & 0.46 & 0.55 & 0.35 & 0.76 & 0.61  \\
        \textbf{Mol-Transformer}  & 0.70 & 0.78 & 0.59 & 0.96 & \textbf{0.70}  \\
        \textbf{Ours}  & \textbf{0.74} & \textbf{0.80} & \textbf{0.62} & \textbf{0.98} & 0.69  \\

        \bottomrule
    \end{tabular}
    \caption{Comparisons with traditional approaches on optimizing molecules' Solubility and BBBP value. We choose success rate, valid rate and similarity as evaluation metrics. The Solubility and BBBP columns display the success rates of the model optimizing these two individual properties respectively, while the "All" column shows the success rate of the model simultaneously optimizing both properties. Our model has achieved the highest success rates in both single-property and multi-property optimization while maintaining high validity and high similarity to the molecules to be optimized. Furthermore, we can also observe that the Transformer architecture performs much better than the Seq2Seq with attention architecture on this task.}
\label{tab:cmp-dmo}
\end{table}

\textbf{Comparisons with LLMs} The results are shown in Table ~\ref{tab:cmp-llms}. Our model significantly outperforms other LLMs in terms of both success rate and valid rate across all tasks. 

From the perspective of the valid ratio of generating valid molecules, BioMedGPT-LM performs poorly. The main reason is that it has difficulty understanding the optimization requirements, often generating content such as guiding users to websites for molecule optimization, rather than outputting the optimized molecule. Although GPT-3.5-turbo appears to have high valid ratio, it often generates molecules that are identical to the given molecule to be optimized, thus failing to serve the purpose of molecule optimization. Our model, on the other hand, demonstrates a significant advantage in generating valid molecules, with virtually no instances of misunderstanding requirements or generating identical molecules to the ones to be optimized.

From the perspective of accuracy, even when we use multi-turn dialogues to prompt the baseline LLMs for comparison, they still struggle to complete the optimization tasks, with low success rates even on relatively simple tasks that only require increasing a single property value.

Our model exhibits good molecule optimization capabilities and strong adaptability to different properties and optimization objectives. Even though our model has only been exposed to data with
individual properties during training, it achieves competitive results in multi-property optimization tasks. However, we also note that our model has a relatively lower success rate in tasks specifying the range of post-optimization property values (``esol+ strict'' task). How to better achieve these more challenging optimization objectives is worth further exploration in the future.

\begin{table}[h]
    \centering
    \begin{tabular}{llcc}
        \toprule
        \textbf{Task} & \textbf{Model} & \textbf{Valid ratio (loose/strict)} & \textbf{Correct ratio (loose/strict)} \\
        \midrule
         \multirow{4}*{\textit{qed+}} & Llama2-7B-Chat & 0.69 / 0.55 & 0.17 
 / 0.16 \\
                    & GPT-3.5-turbo  & 0.97 / 0.96 & 0.15 / 0.15 \\
                    & BioMedGPT-LM  & 0.34 / 0.32 & 0.15 / 0.09 \\
                    & Ours  & \textbf{0.99} / \textbf{0.97} & \textbf{0.76} / \textbf{0.63} \\
        
        \midrule
         \multirow{4}*{\textit{acceptor+}} & Llama2-7B-Chat & 0.45 / 0.43 & 0.08 / 0.08 \\
                    & GPT-3.5-turbo  & \textbf{0.98} / 0.96 & 0.04 / 0.06 \\
                    & BioMedGPT-LM  & 0.45 / 0.39 & 0.18 / 0.13 \\
                    & Ours  & 0.97 / \textbf{0.96} & \textbf{0.71} / \textbf{0.67} \\
        \midrule
         \multirow{4}*{\textit{donor+}} & Llama2-7B-Chat & 0.45 / 0.48 & 0.15 / 0.08 \\
                    & GPT-3.5-turbo  & 0.98 / 0.95 & 0.10 / 0.04 \\
                    & BioMedGPT-LM  & 0.46 / 0.46 & 0.17 / 0.09 \\
                    & Ours  & \textbf{0.98} / \textbf{0.95} & \textbf{0.72} / \textbf{0.76} \\
        \midrule
         \multirow{4}*{\textit{solubility+}} & Llama2-7B-Chat & 0.56 / 0.56 & 0.36 / 0.20 \\
                    & GPT-3.5-turbo  & 0.94 / 0.95 & 0.16 / 0.05 \\
                    & BioMedGPT-LM  & 0.27 / 0.35 & 0.18 / 0.09 \\
                    & Ours  & \textbf{0.98} / \textbf{0.98} & \textbf{0.80} / \textbf{0.41} \\
        \midrule
        \multirow{4}*{\textit{bbbp+}} & Llama2-7B-Chat & 0.56 / 0.57 & 0.19 / 0.14 \\
                    & GPT-3.5-turbo  & 0.97 / 0.95 & 0.10 / 0.10 \\
                    & BioMedGPT-LM  & 0.26 / 0.22 & 0.16 / 0.07 \\
                    & Ours  & \textbf{0.99} / \textbf{0.98} & \textbf{0.82} / \textbf{0.61} \\
        \midrule
       \multirow{4}*{\textit{herg-}} & Llama2-7B-Chat & 0.59 / 0.55 & 0.39 / 0.31 \\
                    & GPT-3.5-turbo  & 0.98 / 0.97 & 0.13 / 0.15 \\
                    & BioMedGPT-LM  & 0.20 / 0.18 & 0.13 / 0.12 \\
                    & Ours  & \textbf{0.99} / \textbf{0.98} & \textbf{0.71} / \textbf{0.67} \\
        \midrule
        \multirow{4}*{\textit{sol+ \& acc+}} & Llama2-7B-Chat & 0.55 / 0.52 & 0.15 / 0.04 \\
                    & GPT-3.5-turbo  & 0.92 / 0.91 & 0.09 / 0.02 \\
                    & BioMedGPT-LM  & 0.29 / 0.32 & 0.10 / 0.07 \\
                    & Ours  & \textbf{0.95} / \textbf{0.95} & \textbf{0.50} / \textbf{0.27} \\
        \midrule
         \multirow{4}*{\textit{qed+ \& bbbp+}} & Llama2-7B-Chat & 0.52 / 0.56 & 0.14 / 0.09 \\
                    & GPT-3.5-turbo  & 0.96 / 0.95 & 0.09 / 0.06 \\
                    & BioMedGPT-LM  & 0.35 / 0.36 & 0.16 / 0.11 \\
                    & Ours  & \textbf{0.99} / \textbf{0.98} & \textbf{0.65} / \textbf{0.41} \\

        \bottomrule
    \end{tabular}
    \caption{Comparisons with LLMs. We evaluated the performance of LLMs on 16 tasks, covering all three optimization objectives introduced in Section 3.1 - loose, strict, and range. We calculated the valid ratio (number of valid SMILES generated/total number in the test set) and success rate (number of molecules meeting optimization objectives/total number in the test set) of the generated molecules. In the task naming, ``+'' represents the goal of increasing the property value, while ``-'' represents property the attribute value. The ``\&'' symbol represents the simultaneous optimization of two given properties. ``sol'' stands for ``Solubility'', and ``acc'' stands for ``the number of hydrogen bond acceptor''. ``loose'' and ``strict'' are the two criteria for defining successful optimization, as detailed in Section 4.1.}
\label{tab:cmp-llms}
\end{table}

\subsection{Case Study}
In this section, we showcase the exceptional capabilities of our model in molecule optimization tasks through several specific examples, beyond its high success rate.

\begin{figure}[h]
    \centering
    \includegraphics[width=1\textwidth]{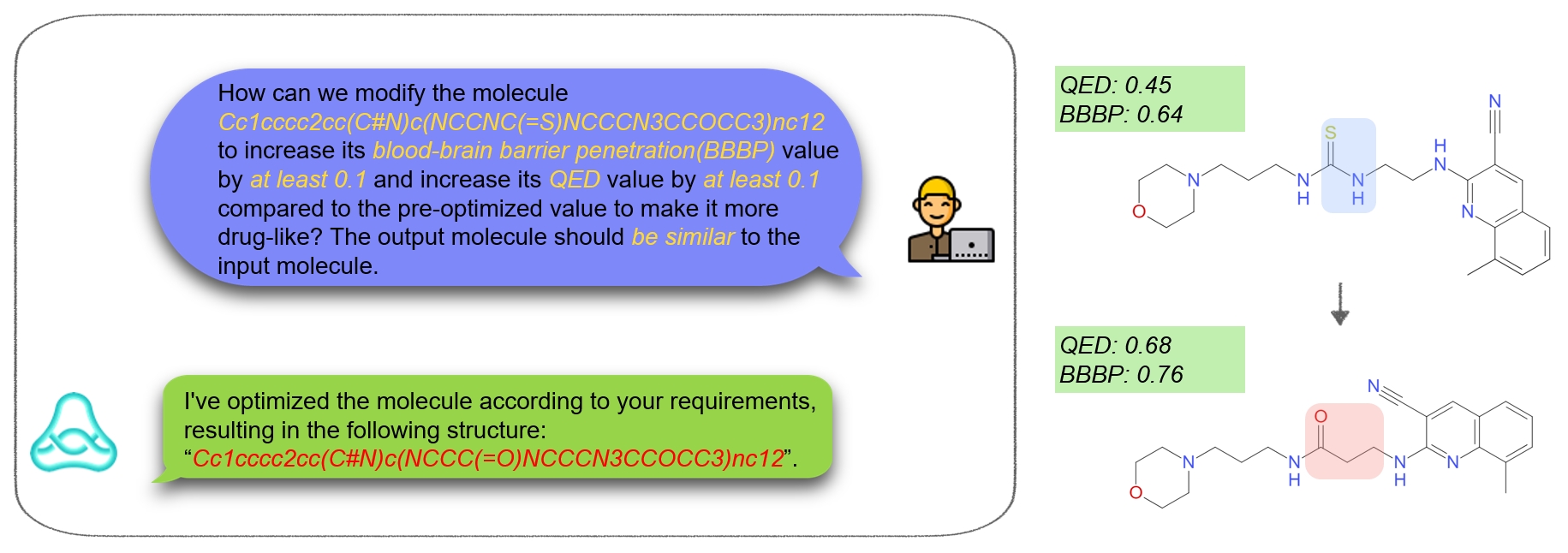}
    \caption{Good transferability of DrugAssist under the zero-shot setting. Users can freely combine individual properties in training data to request DrugAssist to optimize them simultaneously.}
    \label{fig:case_s1}
\end{figure}

\textbf{Transferability} Figure ~\ref{fig:case_s1} demonstrates the good transferability of our model under the \textit{zero-shot} setting. We randomly selected two properties, BBBP and QED, and asked DrugAssist to increase their values by at least 0.1 simultaneously. Our model achieved this and the resulting molecule is structurally similar to the original one. Although the model has only been exposed to data with individual properties during training, users can still freely combine these properties when using the model to optimize them simultaneously. Traditional models, however, often require retraining on new datasets with multiple properties in order to achieve this.

\begin{figure}[h]
    \centering
    \includegraphics[width=1\textwidth]{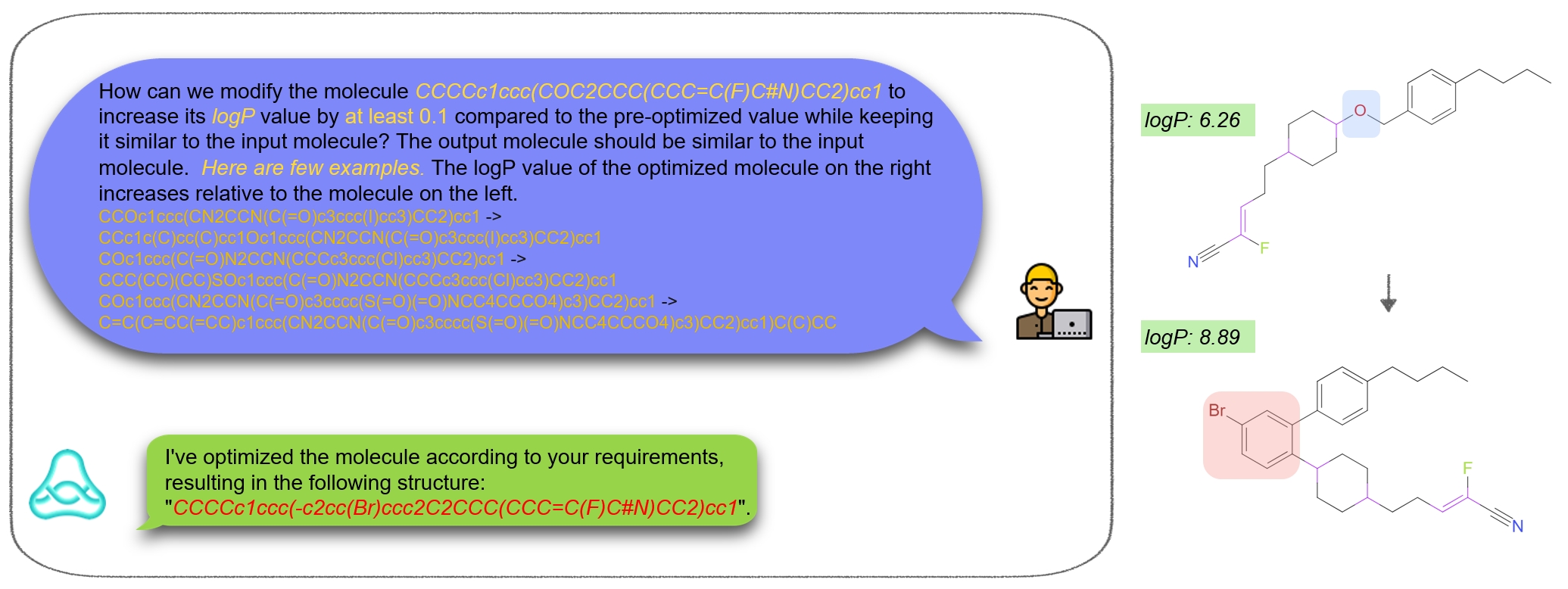}
    \caption{Good transferability of DrugAssist under the few-shot setting. By providing examples of successful optimizations for molecules similar to the one to be optimized, DrugAssist can optimize properties not encountered during training.}
    \label{fig:case_s2}
\end{figure}

Figure ~\ref{fig:case_s2} demonstrates the good transferability of DrugAssist under the \textit{few-shot} setting. We asked DrugAssist to increase the logP value of a given molecule by at least 0.1, even though this property is not included in the training data. By providing a few examples of similar molecules with successfully increased logP values by at least 0.1 in the prompt, our model was able to achieve this. Our model can optimize properties not encountered during training through few-shot, which is difficult to achieve for traditional models (e.g., JT-VAE).

\begin{figure}[h]
    \centering
    \includegraphics[width=1\textwidth]{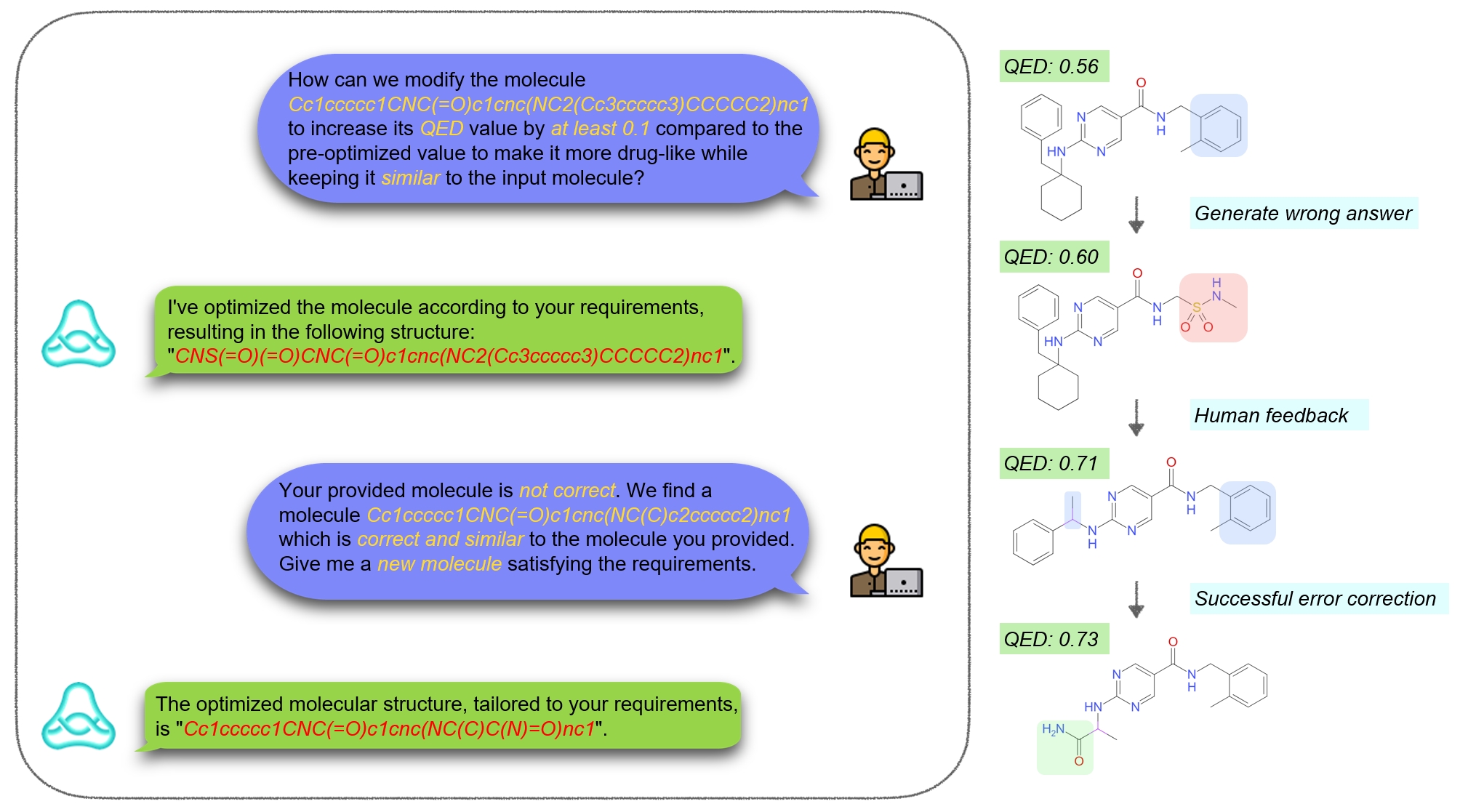}
    \caption{Iterative optimization capability of DrugAssist. when the model provides a molecule that does not fully meet the requirements, it can correct the error and generate a new, compliant molecule based on a human-provided example.}
    \label{fig:case_s3}
\end{figure}

\textbf{Iterative optimization} Figure ~\ref{fig:case_s3} illustrates the iterative optimization capability of our model. We asked DrugAssist to increase the QED value of a given molecule by at least 0.1, but it failed. Then, we found a molecule from the database that was similar to the failed molecule it provided and met the optimization requirements as a hint for it to generate a new one. This time, it chose to optimize different functional groups than the first time and succeeded, and the final molecule generated is structurally similar to the original given molecule. We can conclude that when the model provides a molecule that does not fully meet the requirements, it can correct the error and generate a new, compliant molecule based on a human-provided example that meets the criteria. This ability highlights the potential for DrugAssist to assist researchers in continually adjusting and optimizing molecules in real-world scenarios.

\section{Conclusion and Future Work}

In this paper, we present DrugAssist, an interactive molecule optimization model. Unlike previous methods, DrugAssist can interact with humans in real-time using natural language. It can provide optimized results based on the instructions given by users and continue to adjust according to their feedback. It demonstrates excellent performance in both single-property and multi-property optimization, including more challenging tasks, such as optimizing within specified property value ranges. Additionally, it shows great potential in transferability and iterative optimization capabilities during the interaction process. Furthermore, we publicly release MolOpt-Instructions, an instruction-based dataset to facilitate future work on fine-tuning LLMs in the molecule optimization domain.

In the future, we aim to improve the model's ability to handle multimodal data and tasks \citep{lyu2023macaw,li2023comprehensive-1} to reduce hallucination problems \citep{zhang2023siren,liu2023retrieval,li2023comprehensive-2,cai2023comprehensive}. Additionally, we are endeavoring to further enhance DrugAssist's interactive capabilities to better understand users' needs and feedback.



\bibliography{iclr2024_conference}
\bibliographystyle{iclr2024_conference}

\clearpage

\appendix
\section{Prompt Settings for the comparisons with LLMs}

\begin{table}[h]
\centering
\begin{tabular}{@{}lp{11cm}@{}}
\toprule
\textbf{Task} & \textbf{Prompt} \\
\midrule
qed+ loose & Help me make molecule [SMILES] more like a drug. The output molecule should be similar to the input molecule. \\
\midrule
qed+ strict & How can we modify the molecule [SMILES] to increase its QED value by at least 0.1 compared to the pre-optimized value to make it more drug-like while keeping it similar to the input molecule?\\
\midrule
acceptor+ loose & Can you make molecule [SMILES] with more hydrogen bond acceptors? The output molecule should be similar to the input molecule.\\
\midrule
acceptor+ strict & Help me increase the number of hydrogen bond acceptors in the molecule [SMILES] by at least 2 compared to the pre-optimized value. The output molecule should be similar to the input molecule.\\
\midrule
donor+ loose & Can you make molecule [SMILES] with more hydrogen bond donors? The output molecule should be similar to the input molecule.\\
\midrule
donor+ strict & Help me increase the number of hydrogen bond donors in the molecule [SMILES] by at least 2 compared to the pre-optimized value. The output molecule should be similar to the input molecule.\\
\midrule
solubility+ loose & How can we modify the molecule [SMILES] to increase its water solubility value while keeping it similar to the input molecule?\\
\midrule
solubility+ strict & Can you give me an optimized version of the molecule [SMILES] with a water solubility value ranging from \textit{lower bound} to \textit{higher bound} (logarithm of mol/L) while maintaining similarity to the original molecule?\\
\midrule
bbbp+ loose & How can we modify the molecule [SMILES] to increase its blood-brain barrier penetration (BBBP) value while keeping it similar to the input molecule?\\
\midrule
bbbp+ strict & How can we modify the molecule [SMILES] to increase its blood-brain barrier penetration (BBBP) value by at least 0.1 compared to the pre-optimized value while keeping it similar to the input molecule?\\
\midrule
herg- loose & How can we modify the molecule [SMILES] to decrease its hERG inhibition value while keeping it similar to the input molecule?\\
\midrule
herg- strict & How can we modify the molecule [SMILES] to decrease its hERG inhibition value by at least 0.1 compared to the pre-optimized value while keeping it similar to the input molecule?\\
\midrule
sol+ \& acc+ loose & How can we modify the molecule [SMILES] to increase its water solubility value and to have more hydrogen bond acceptors? The output molecule should be similar to the input molecule.\\
\midrule
sol+ \& acc+ strict & Can you give me an optimized version of the molecule [SMILES] with a water solubility value ranging from \textit{lower bound} to \textit{higher bound} (logarithm of mol/L), and with at least 2 more hydrogen bond acceptors while maintaining similarity to the original molecule?\\
\midrule
qed+ \& bbbp+ loose & How can we modify the molecule [SMILES] to increase its blood-brain barrier penetration (BBBP) value and make it more like a drug? The output molecule should be similar to the input molecule.\\
\midrule
qed+ \& bbbp+ strict & How can we modify the molecule [SMILES] to increase its blood-brain barrier penetration (BBBP) value by at least 0.1 and increase its QED value by at least 0.1 compared to the pre-optimized value to make it more drug-like? The output molecule should be similar to the input molecule.\\

\bottomrule
\end{tabular}
\caption{Prompts for different tasks. ``[SMILES]'' represents the SMILES string for the molecule.}
\label{tab:prompt_chatdrug}
\end{table}

\end{document}

%% file: math_commands.tex
\usepackage{amsmath,amsfonts,bm}

\def\eqref#1{equation~\ref{#1}}

\def\1{\bm{1}}

\DeclareMathAlphabet{\mathsfit}{\encodingdefault}{\sfdefault}{m}{sl}
\SetMathAlphabet{\mathsfit}{bold}{\encodingdefault}{\sfdefault}{bx}{n}

